\documentclass[journal]{IEEEtran} 

\usepackage{tabularx}
\usepackage{graphicx}  
\usepackage{amsfonts}
\usepackage{amssymb}
\usepackage{amsmath}
\usepackage{mathrsfs}
\usepackage{multirow}
\usepackage{psfrag}
\usepackage{color}
\usepackage{amsthm}
\usepackage{bbold}

\usepackage{tikz}
\usepackage{textcomp}
\usepackage{hyperref}
\usepackage{lipsum}

\newtheorem{the:def}{Definition}	 	
\newtheorem{the:lem}[the:def]{Lemma}	
\newtheorem{the:the}[the:def]{Theorem}	
\newtheorem{the:rem}[the:def]{Remark}	  
\newtheorem{the:cor}[the:def]{Corollary}
\newtheorem{the:alg}[the:def]{Algorithm}
\newtheorem{the:pro}[the:def]{Problem}
\newtheorem{the:ass}[the:def]{Assumption}

\title{\LARGE \bf Consensus analysis of large-scale nonlinear homogeneous  multi-agent formations with  polynomial dynamics}

\author{Paolo Massioni, G\'erard Scorletti
\thanks{P. Massioni is with Laboratoire Amp\`ere, UMR CNRS 5005, INSA de Lyon, Universit\'e de Lyon, F-69621 Villeurbanne, France
  {\tt\small  paolo.massioni@insa-lyon.fr}}
  \thanks{G. Scorletti is with Laboratoire Amp\`ere, UMR CNRS 5005, Ecole Centrale de Lyon, Universit\'e de Lyon, F-69134 Ecully, France
  {\tt\small  gerard.scorletti@ec-lyon.fr}}
}

\begin{document}

\maketitle
\thispagestyle{empty}

\begin{abstract}
Drawing inspiration from the theory of linear ``decomposable systems'', we provide a method, based on linear matrix inequalities (LMIs), which makes it possible to prove the convergence (or consensus) of a set of interacting agents with polynomial dynamic. We also show that the use of a generalised version of the famous Kalman-Yakubovic-Popov lemma allows the development of an LMI test whose size does not depend on the number of agents.
The method is validated experimentally on two academic examples. 
\end{abstract}

%

\section{Introduction}

Large-scale systems are an emerging topic in the system and control community, which is devoting significant efforts on the development of analysis and control synthesis methods for them. This deep interest can clearly be seen from the large number of works published in the field in the last 40 years \cite{bamiehmain,dandrea,davison1979sequential,scorlettidec,faxmurray,langbortgraph,massioni2008distributed,popov2009,li2010consensus,massioni2014distributed}. 

One of the main objectives of the studies is the development and validation of ``distributed control laws'' for obtaining a certain specified goal for a system of this kind.
By ``distributed control'', opposed to ``centralized control'', we mean a control action that is computed locally according to the physical spatial extension of the system, which is seen as an interconnection of simpler subsystems. One of the main problems of large-scale systems is the ``curse of dimensionality'' that goes with them, i.e. the analysis and synthesis problems related to dynamical systems grow with the size, and for system of very high order, such problems becomes computationally infeasible.
In the literature, if we restrict to linear systems, we can find a few solutions  \cite{massioni2008distributed,faxmurray,bamiehmain} that can effectively overcome the curse of dimensionality for a class of systems with a certain regularity, namely for what we call ``homogeneous systems'' which are made of the interconnection of a huge number of identical subunits (also sometimes called  ``agents'').

In this paper we focus on this same problem, more specifically on the stability analysis of large-scale dynamical systems, for the nonlinear case. Namely, we will consider a formation or a multi-agent system made of a high number of identical units interacting with one another through a symmetric graph-like interconnection. Assuming that the dynamical equation of each subunit is described by a polynomial in the state vector, we can show that a linear matrix inequality (LMI) test can be devised in order to verify the relative stability of such a formation. We will also be able to formulate such a test in a form which is not strictly depending on the formation size, making it possible to check the stability of formations with virtually any number of agents, basically extending the analysis results of \cite{massioni2008distributed,ghadami2013decomposition} to the nonlinear (polynomial) case.

\section{Preliminaries}
\label{sec:pre}

\subsection{Notation}
We denote by \(\mathbb{N}\) the set of natural numbers,  by \(\mathbb{R}\) the set of real numbers and by \(\mathbb{R}^{n \times m}\) the set of real  \({n \times m}\) matrices. 
 \(A^\top\) indicates the transpose of a matrix \(A\),  \(I_n\) is the identity matrix of size \(n\),   \(\mathbf{0}_{n \times m}\) is a matrix of zeros of size \(n \times m\) and \(\mathbf{1}_n \in \mathbb{R}^n\) a column vector that contains \(1\) in all of its entries. The notation \(A \succeq 0\) (resp. \(A \preceq 0\)) indicates that all the eigenvalues of the symmetric matrix \(A\) are positive (resp. negative) or equal to zero, whereas \(A \succ 0\) (resp. \(A \prec 0\)) indicates that all such eigenvalues are strictly positive (resp. negative). The symbol \(\dbinom{n}{k}\) indicates the binomial coefficient, for which we have \[\dbinom{n}{k}=\frac{n!}{k!(n-k)!}.\]
The symbol \(\otimes\) indicates the Kronecker product, for which we remind the basic properties \((A\otimes B)(C \otimes D)=(AC\otimes BD)\), \((A \otimes B)^\top=(A^\top \otimes B^\top)\), \((A \otimes B)^{-1}=(A^{-1} \otimes B^{-1})\)  (with matrices of compatible sizes).
We employ the symbol \(\ast\) to complete symmetric matrix expressions avoiding repetitions.

\subsection{Agent dynamic}

We consider a set of \(N \in \mathbb{N}\) identical agents or subsystems of order \(n\), which interact with one another.
Each agent, if taken alone, is supposed to be described by a polynomial dynamic, of the kind
\begin{equation}
\dot{x}_i=f_d(x_i) = A_a \chi_i
\label{eq:agent}
\end{equation}
where \(i=1,...,N\), \(x_i=[ x_{i,1}, \,x_{i,2},\,...,\,x_{i,n} ]^\top \in \mathbb{R}^{n}\) is the state of the \(i\)\textsuperscript{th} agent, \(f_d\) is a polynomial function of degree  \(d\in \mathbb{N}\), \(A_a \in \mathbb{R}^{n \times \rho}\) and \( \chi_i \in \mathbb{R}^\rho\) is the vector containing all the monomials in \(x_i\) up to degree \(d\) 
(for example, if \(n=2\), \(d=2\), then \(\chi_i=[1, \, x_{i,1}, \,x_{i,2}, \, x_{i,1}^2,\, x_{i,1} x_{i,2}, \, x_{i,2}^2 ]^\top\)). The value of \(\rho\) is given by
\begin{equation}
\rho = \dbinom{n+d}{n}.
\label{eq:rho}
\end{equation}
This approach is based on the sum of squares (SOS) literature  \cite{parrilo2003semidefinite}, which basically allows the relaxation of polynomial problems into linear algebra's. In this context, it is  possible to express polynomials \(p\) up to degree \(2d\) as quadratic forms with respect to \(\chi_i\), i.e. \(p(x_i)=\chi_i^\top  \mathcal{X} \chi_i\), with \(\mathcal{X}=\mathcal{X}^\top  \in \mathbb{R}^{\rho \times \rho} \). This quadratic expression is not unique, due to the fact that different products of monomials in \(\chi_i\) can yield the same result, for example \(x_{i,1}^2\) is either \(x_{i,1}^2\) times \(1\) or \(x_{i,1}\) times \(x_{i,1}\). This implies that there exist linearly independent slack matrices \(Q_k=Q_k^\top  \in \mathbb{R}^{\rho \times \rho}\), with \(k=1,\ldots,\iota\) such that \(\chi_i^\top  Q_k \chi_i =0\). The number of such matrices is \cite{parrilo2003semidefinite}
\begin{equation}
\iota=\frac{1}{2}\left( {\dbinom{d+n}{d}}^2 +   \dbinom{d+n}{d} \right) -       \dbinom{n+2d}{2d}.
\label{eq:iota}
\end{equation}

\subsection{Formations}
Moving from one single agent to a whole formation, we employ a matrix \(\mathcal{P} \in \mathbb{R}^{N \times N}\) to describe the interactions among the agents. Basically \(\mathcal{P}\) is a sparse matrix whose entries in the \(i\)\textsuperscript{th} row and \(j\)\textsuperscript{th} column indicate whether the \(i\)\textsuperscript{th} agent is influenced by the state of the \(j\)\textsuperscript{th}, according to the definition that follows.

\begin{the:def}[Formation] We call a formation (of non-linear agents with polynomial dynamics) a dynamical system of order \(nN\), with \(n,N \in \mathbb{N}\), described by the following dynamical equation
\begin{equation}
\dot{x} = (I_N \otimes A_a  + \mathcal{P} \otimes A_b) \chi 
\label{eq:formation}
\end{equation}
where 
\(x=[ x_1^\top,\, x_2^\top,\, ...,\,x_N^\top ]^\top \in \mathbb{R}^{nN}\),
\(\chi=[ \chi_1^\top,\, \chi_2^\top,\, ...,\,\chi_N^\top ]^\top \in \mathbb{R}^{\rho N}\),
\(\mathcal{P} \in \mathbb{R}^{N \times N}\)
and 
\(A_a, A_b \in \mathbb{R}^{n \times \rho}\).
\end{the:def}

This definition extends and adapts the definition of ``decomposable systems'' given in \cite{massioni2008distributed} to  polynomial dynamics. In the linear case, a formation defined above boils down to the dynamical equation
\begin{equation}
\dot{x} = (I_N \otimes A_a  + \mathcal{P} \otimes A_b) x. 
\label{eq:formation_linear}
\end{equation}
In \cite{massioni2008distributed} it has been shown that if \(\mathcal{P}\) is diagonalisable, then this system (of order \(nN\)) is equivalent to a set of parameter-dependent linear systems of order \(n\). This is obtained with the change of variables \(x=(S \otimes I_n)\hat{x}\), where 
\(\hat{x}=[ \hat{x}_1^\top,\, \hat{x}_2^\top,\, ...,\,\hat{x}_N^\top ]^\top \in \mathbb{R}^{nN}\) and \(S\) is the matrix diagonalising \(\mathcal{P}\), i.e. \(S^{-1}\mathcal{P}S = \Lambda\), with \(\Lambda\) diagonal. This turns \eqref{eq:formation_linear} into \(\dot{\hat{x}} = (I_N \otimes A_a + \Lambda \otimes A_b)\hat{x}\), which is a block-diagonal system equivalent to the set 
\begin{equation}
\dot{\hat{x}}_i =  (A_a  + \lambda_i A_b) \hat{x}_i 
\label{eq:formation_linear_decomposed}
\end{equation}
for \(i=1,...,N\), with \(\lambda_i\) the \(i\)\textsuperscript{th} eigenvalue of \(\mathcal{P}\).
This idea of decomposing a distributed system into a set of parameter-varying systems is very practical and it has inspired several works in the domain of consensus and distributed control \cite{li2011h,zakwan2016polynomial,eichler2014robust,demir2011decomposition}. In this paper, to our knowledge, the idea is adapted to nonlinear dynamics for the first time.

\subsection{Problem formulation}
The topic of this paper is to find a proof of convergence of the state of the agents under a given dynamics expressed as in \eqref{eq:formation}. We do not require that each agent by itself converges to a point, but that they all converge eventually to the same state, which could be either an equilibrium point or a trajectory. In order to do so, we formulate first an assumption on the pattern matrix \(\mathcal{P}\).
\begin{the:ass}
The pattern matrix \(\mathcal{P} \in \mathbb{R}^{N \otimes N}\) in \eqref{eq:formation} is symmetric and it has one and only one eigenvalue equal to \(0\), associated to the eigenvector \(\mathbf{1}_N \),   i.e \(\mathcal{P} \mathbf{1}_N=0\). 
\label{ass:P}
\end{the:ass}

This assumption is very common in the literature, it basically ensures that the interconnection matrix is a (generalised) graph Laplacian of a symmetric connected graph  \cite{graphtheorybook}. Such matrices have real eigenvalues and eigenvectors.
We can then formulate the problem on which this paper focuses.

\begin{the:pro}
We consider \eqref{eq:formation} with initial conditions \(x(0) \in \mathbb{R}^{nN}\). Prove that \(\lim_{t\to\infty} ||x_i-x_j|| = 0\), \(\forall i,j \in \{1,\,...,\,N \} \). 
\label{pro:pro}
\end{the:pro}
  
\section{Formation Lyapunov function}
\label{sec:lya}

In order to be able to prove the convergence of all the agents to the same trajectory, we define what we call a ``formation Lyapunov function'', which will have the property of tending to zero when the agents are converging. We summarise these notions in a definition and a lemma.

\begin{the:def}[Formation Lyapunov function candidate] We define as ``formation Lyapunov function candidate'' a function 
\begin{equation}
V(x) =x^\top \left(\sum_{i=1}^l \mathcal{P}^i \otimes L_i\right) x=x^\top \mathcal{L} x
\label{eq:flf}
\end{equation}
with \(L_i=L_i^\top \in \mathbb{R}^{n \times n}\), \(l \in \mathbb{N}\), \(l \leqslant N\).
\label{def:flf}
The reason for this special structure will be clear later on, in fact it allows the block-diagonalisation of the Lyapunov matrix in the same way as \(\mathcal{P}\) can be diagonalised.

\end{the:def}

\begin{the:lem}
Consider \eqref{eq:formation} and a  formation Lyapunov function candidate \(V(x)=x^\top \mathcal{L} x \) as in \eqref{eq:flf}. Let \(\mathbf{1}_N^\bot \in \mathbb{R}^{N\times (N-1) }\) be the orthogonal complement of \(\mathbf{1}_N\), i.e. \([\mathbf{1}_N \,\, \mathbf{1}_N^\bot ]\) is full rank and \(\mathbf{1}_N^\top \mathbf{1}_N^\bot  =0\).

If \((\mathbf{1}_N^\bot \otimes I_n)^\top \mathcal{L}\, (\mathbf{1}_N^\bot  \otimes I_n) >0\), then we have that \(x_i = x_j\) \(\forall i,j \in \{1,\,...,\,N \} \) if and only if \(V(x)=0\).
\begin{proof}
Necessity is almost obvious: if \(x_i = x_j\) \(\forall i,j \in \{1,\,...,\,N \} \), then \(x=\mathbf{1}_N  \otimes x_i\); the fact that \(\mathcal{P} \mathbf{1}_N  =0\) implies that \(V(x)=0\).

We prove the sufficiency by contradiction, i.e. we suppose that there exist \(i\) and \(j\) for which \(x_i \neq x_j\) and \(V(x)=0\). The vector \(x\) with the complete state must then have at least one component which is orthogonal to the columns of \( (\mathbf{1}_N  \otimes I_n)\), because  \((\mathbf{1}_N  \otimes I_n)\) contains columns with all the corresponding agent states equal. So, based on the fact that  \((\mathbf{1}_N^\bot \otimes I_n)^\top \mathcal{L}\, (\mathbf{1}_N^\bot  \otimes I_n) >0\), then \(V(x)>0\)
contradicting the hypothesis.
\end{proof}
\label{lem:con}
\end{the:lem}

\section{Main result}
\label{sec:mai}

We are now ready to formulate our main result. A preliminary lemma is given first, which allows diagonalising the Lyapunov function in the same way as a linear system is decomposed in \cite{massioni2008distributed}.

\begin{the:lem}
\label{lem:S}
If Assumption~\ref{ass:P} holds, then 1) there exist a matrix \(S \in \mathbb{R}^{N\times N}\) such that \(S^\top S = S S^\top = I_N\), and \(S^\top \mathcal{P}S = \Lambda\), with \(\Lambda\) diagonal. Moreover, we have that 2) \(S^\top  \mathbf{1}_N = T = [t_1 \,\, t_2 \,\,...\,\, t_N]\), with \(t_i \in \mathbb{R}^{N} = 0 \) if \(\lambda_i=\Lambda_{i,i} \neq 0\).
\begin{proof}
The first part of the lemma is proven by the fact that all symmetric matrices are diagonalisable by an orthonormal matrix \(S\) (i.e. \(S^{-1}=S^\top\)) \cite{matrixcomputations}. For the second part, consider that due to Assumption~\ref{ass:P}, \(\mathbf{1}_N \) is an eigenvector of \(\mathcal{P}\) with eigenvalue \(0\); the matrix \(S\) contains the normalised eigenvectors of \(\mathcal{P}\) in its columns, and all of these eigenvectors are orthogonal to one another because \(S^\top S = I_N\). So each \(t_i\) is the dot product between \(\mathbf{1}_N \) and the \(i\)\textsuperscript{th} eigenvector, and it is non zero if and only if \(\lambda_i=0\).
\end{proof}
\end{the:lem}

\begin{the:the}
Consider \eqref{eq:formation} with given \(N\), \(A_a\), \(A_b\) and \(\mathcal{P}\) statisfying Assumption~\ref{ass:P}; moreover, we order the eigenvalues of \(\mathcal{P}\) so that the first eigenvalue is the one equal to zero, i.e. \(\lambda_1=0\).
If for a chosen \(l \in \mathbb{N}\), there exist \(\tau_j \in \mathbb{R}\), and  matrices \(L_j=L_j^\top \in \mathbb{R}^{n \times n } \) such that
\begin{equation}\begin{array}{c}
  \sum_{j=1}^l \lambda_i^j  L_j 
 \succ 0
 \end{array}
\label{eq:the1.1}
\end{equation}
\begin{equation}\begin{array}{c}\Pi (\sum_{j=1}^{\iota}{\tau_j Q_j}+  \sum_{j=1}^l (\lambda_i^j (\Gamma^\top L_j  A_a + A_a^\top L_j \Gamma + \Gamma^\top L_j \Gamma ) +\\
  \lambda_i^{j+1} (\Gamma^\top L_j  A_b + A_b^\top L_j \Gamma)) \Pi^\top
 \preceq 0
 \end{array}
\label{eq:the1.2}
\end{equation}
for \(i=2,\,...\,N\), where \(\Gamma= [ \mathbf{0}_{n,1} \,\, I_{n}\,\, \mathbf{0}_{n,\rho-n-2} ]\) (i.e. \(\Gamma x_i = \chi_i \)), \(\Pi = [\mathbf{0}_{\rho-1,1}\,\,I_{\rho-1}  ] \),
then
 \(\lim_{t\to\infty}|| x_i-x_j|| = 0\), \(\forall i,j \in \{1,\,...,\,N \} \).

\begin{proof}
In order to assure the convergence of the agents, we need to assure the conditions stated in Lemma~\ref{lem:con}, namely that a function \(V(x)= x^\top \mathcal{L} x \) exists, with and \((\mathbf{1}_N^\bot \otimes I_n)^\top \mathcal{L}\, (\mathbf{1}_N^\bot  \otimes I_n) >0\), and that
 \(\dot{V}(x)<0\) for \(V(x)> 0\).
 For the condition \((\mathbf{1}_N^\bot \otimes I_n)^\top \mathcal{L}\, (\mathbf{1}_N^\bot  \otimes I_n) >0\), consider that \(S\) contains a scaled version of \(\mathbf{1}_N\) in its first column and \(\mathbf{1}_N^\bot\) in the rest of the matrix. Knowing that \(\mathcal{L}=(S \otimes I_n)(\sum_{i=1}^l \Lambda^i\otimes L_i)(S^\top \otimes I_n)\) thanks to Lemma~\ref{lem:S}, this condition is equivalent to \eqref{eq:the1.1}. 
 
 For what concerns the condition \(\dot{V}(x)<0\) for \(V(x) > 0\), it is satisfied if \(\dot{V}(x)\leqslant -\epsilon x^\top \mathcal{L} x\), i.e.
\begin{equation}
\begin{array}{c}\chi ^\top (\mathcal{Q}_N+\Gamma_N^\top \mathcal{L}(I_N \otimes A_a)  +\Gamma_N^\top \mathcal{L}(\mathcal{P} \otimes A_b) +\\
  (I_N \otimes A_a^\top)\mathcal{L}\Gamma_N  + (\mathcal{P} \otimes A_b^\top)\mathcal{L} \Gamma_N + \epsilon  \Gamma_N^\top \mathcal{L}  \Gamma_N )\chi \leqslant 0
 \end{array}
\label{eq:the10}
\end{equation}
where \(\Gamma_N = (I_N \otimes \Gamma)\) (so \(\Gamma_N \chi=x\)) and \(\mathcal{Q}_N = I_N \otimes \sum_{j=1}^{\iota}{\tau_j Q_j}\) (for which, by definition of \(Q_j\), \(\chi^\top \mathcal{Q}_N \chi = 0\) for all values of the \(\tau_i\)). 
By the fact that \(\mathcal{P}= S \Lambda S^\top  \) and   \({I}_N= S  S^\top  \) (Assumption~\ref{ass:P} and Lemma~\ref{lem:S}), using the properties of Kronecker product \eqref{eq:the10} is equivalent to
\begin{equation}
\begin{array}{c}\hat{\chi}^\top (\mathcal{Q}_N+\Gamma_N^\top \hat{\mathcal{L}}(I_N \otimes A_a)  +\Gamma_N^\top \hat{\mathcal{L}}(\Lambda \otimes A_b) +\\
  (I_N \otimes A_a^\top)\hat{\mathcal{L}}\Gamma_N  + (\Lambda \otimes A_b^\top) \hat{\mathcal{L}}\Gamma_N + \epsilon  \Gamma_N^\top \hat{\mathcal{L}}  \Gamma_N )\hat{\chi} \leqslant 0
 \end{array}
\label{eq:the20}
\end{equation}
with \(\hat{\mathcal{L}}=  \sum_{j=1}^{l}{\Lambda^j \otimes L_j}\) and \(\hat{\chi}= (S^\top \otimes I_\rho ) \chi\). Notice in this last inequality that the term between \(\hat{\chi}^\top\) and \(\hat{\chi}\) is block-diagonal, as it is the sum of terms of the kind \(I_N \otimes X\) or \(\Lambda^i \otimes X\) (\(i \in \mathbb{N}\)). 
If we define \(\hat{\chi}_i \in \mathbb{R}^{\rho}\) such that \(\hat{\chi}=[ \hat{\chi}_1^\top,\, \hat{\chi}_2^\top,\, ...,\,\hat{\chi}_N^\top ]^\top \), then \eqref{eq:the20} is equivalent to
\begin{equation}
\begin{array}{c}\sum_{i=1}^{N} \hat{\chi}_i^\top (\sum_{j=1}^{\iota}{\tau_j Q_j}+   \sum_{j=1}^l \lambda_i^j (\Gamma^\top L_j  A_a + A_a^\top L_j \Gamma) +\\
 \!\sum_{j=1}^l \lambda_i^{j+1} \!(\Gamma^\top L_j  A_b + A_b^\top L_j \Gamma) \!
+\!
  \epsilon\sum_{j=1}^l \lambda_i^j (  \Gamma^\top L_j \Gamma ))\hat{\chi}_i\! \leqslant\! 0.
 \end{array}
\label{eq:the30}
\end{equation}
The term of the sum for \(i=1\) is always \(0\) (as we chose \(\lambda_1=0\)), so there is no contribution from it and it can be discarded. Concerning the vectors \(\chi_i\), remember that they all contain \(1\) in their first entry, i.e. \(\chi_i=[1\,\, \tilde{\chi}_i^\top ]^\top\), \(\tilde{\chi}_i \in \mathbb{R}^{\rho-1}\). For each \(\hat{\chi}_i\), the first entry is by its definition the \(i\)\textsuperscript{th} entry of the vector \(e = S^\top \mathbf{1}\), which contains zeros in all of its entries but the first (due to Lemma~\ref{lem:S}). So for \(i=2,\,...,\,N\), we have that \(\hat{\chi}_i=\Pi^\top \Pi\hat{\chi}_i\). So
 \eqref{eq:the30} is equivalent to
\begin{equation}
\begin{array}{c}\sum_{i=2}^{N} \hat{\chi}_i^\top \Pi^\top \Pi (\sum_{j=1}^{\iota}{\tau_j Q_j}+  \\ \sum_{j=1}^l \lambda_i^j (\Gamma^\top L_j  A_a + A_a^\top L_j \Gamma) +\\
 \sum_{j=1}^l \lambda_i^{j+1} (\Gamma^\top L_j  A_b + A_b^\top L_j \Gamma) 
+\\
  \epsilon\sum_{j=1}^l \lambda_i^j (  \Gamma^\top L_j \Gamma ))\Pi^\top \Pi \hat{\chi}_i \leqslant 0
 \end{array}
\label{eq:the40}
\end{equation}
The set of LMIs in \eqref{eq:the1.2} imply \eqref{eq:the40}, which concludes the proof.
\end{proof}
\label{the:mai}
\end{the:the}

This theorem allows proving the convergence of \(N\) agents with two sets of \(N-1\) parameter-dependent LMIs, whose matrix size is respectively \(n\) (i.e. the order of each agent taken alone) and \(\rho-1\). 
This result is already interesting as it avoids using LMIs scaling with \(Nn\), which is the global system order. In the next section, we explore whether it is possible to further reduce the computational complexity.

\section{Variation on the main result}
\label{sec:kyp}

We explore the possibility of using a generalised version of the famous Kalman-Yakubovich-Popov (KYP) lemma \cite{rantzer1996kalman}. This lemma allows turning a parameter-depending LMI into a parameter-independent one.

\subsection{The Kalman-Yakubovic-Popov lemma}

The Kalman-Yakubovic-Popov lemma or KYP \cite{rantzer1996kalman} is a widely celebrated result for dynamical systems that allows turning frequency-dependent inequalities into frequency-independent ones, by exploiting a state-space formulation. It turns out that such a result can be adapted and generalised to inequalities depending on any scalar parameter. Namely, we will use the following generalised version of the KYP.

\begin{the:lem}[Generalized KYP \cite{dinh2005parameter}] Consider 
\begin{equation}
M(\xi) = M_{0} + \sum^{\eta}_{i=1} \xi_{i}M_{i},
\label{eq:kypform}
\end{equation} 
with \( \xi \in \mathbb{R}^{l} \) a vector of decision variables and \( M_{i}= M_{i}^\top  \in \mathbb{R}^{n_M \times n_M}, i=1,...,\eta \).
The quadratic constraint
\begin{equation}
 {\phi (\theta)}^\top  M(\xi)\,\,\phi (\theta)\prec 0 \textrm{ for } \theta \in [\underline{\theta},\overline{\theta}]
\label{eq:phiMphi}
\end{equation}
is verified if and only if there exist  \(\mathcal{D} = \mathcal{D}^\top  \succ 0\) and \(\mathcal{G} = -\mathcal{G}^\top  \) such that
\begin{equation}
\begin{array}{c}
\left[\begin{array}{c} \tilde{C}^\top  \\ \tilde{D}^\top
\end{array}\right]M(\xi)
\left[\begin{array}{cc} \tilde{C} & \tilde{D}
\end{array}\right]+\\
\left[\!\!\begin{array}{ccc} I & 0 \\ \tilde{A} & \tilde{B}
\end{array}\!\!\right]^\top \!\!
\left[\!\!\begin{array}{cccc} -2\mathcal{D} & (\underline{\theta}+\overline{\theta})\mathcal{D}+\mathcal{G} \\ (\underline{\theta}+\overline{\theta})\mathcal{D}-\mathcal{G} & -2\underline{\theta}\overline{\theta}\mathcal{D}
\end{array}\!\!\right] \!\!
\left[\!\!\begin{array}{ccccc} I & 0 \\ \tilde{A} & \tilde{B}
\end{array}\!\!\right]\!\!\prec \!0
\label{eq:kyp1}
\end{array}
\end{equation}
with \( \tilde{A}\) ,\(\tilde{B}\), \(\tilde{C} \) and \( \tilde{D} \) such that
\begin{equation}
\phi (\theta) = \tilde{D}+\tilde{C} \theta I (I-\tilde{A}\theta I)^{-1}\tilde{B} = \theta I \star
\left[
\begin{array}{cc}
\tilde{A} & \tilde{B} \\ \tilde{C} & \tilde{D}
\end{array}
\right],
\label{eq:2.22}
\end{equation}
where the operator \(\star\) implicitly defined above is known as the Redheffer product \cite{zhou1996robust}.

The lemma applies as well if the sign \(\prec\) in \eqref{eq:phiMphi} is replaced by \(\preceq\): in this case replace \(\prec\) with  \(\preceq\) in \eqref{eq:kyp1} as well.
\label{lem:kyp}
\end{the:lem}

\subsection{Second main result}

Let us define
\begin{equation}
\overline{\lambda}=\max_{2 \leqslant i \leqslant N}\{\lambda_i\},\,\,\
\underline{\lambda}=\min_{2 \leqslant i \leqslant N}\{\lambda_i\}.
\label{eq:minmax}
\end{equation}

Then, for \(\theta \in [\underline{\lambda},\,\overline{\lambda}] \), the following set of LMIs 
\begin{equation}\begin{array}{c}
  \sum_{j=1}^l \theta^j  L_j 
 \succ 0
 \end{array}
\label{eq:the1.1b}
\end{equation}
\begin{equation}\begin{array}{c}\Pi (\sum_{j=1}^{\iota}{\tau_j Q_j}+  \sum_{j=1}^l (\theta^j (\Gamma^\top L_j  A_a + A_a^\top L_j \Gamma + \Gamma^\top L_j \Gamma ) +\\
  \theta^{j+1} (\Gamma^\top L_j  A_b + A_b^\top L_j \Gamma)) \Pi^\top
 \preceq 0
 \end{array}
\label{eq:the1.2b}
\end{equation}
 ``embeds'' the set of LMIs in \eqref{eq:the1.1} and \eqref{eq:the1.2} (notice that we have moved from a discrete set of values to a continuous interval which includes them all). 
Subsequently, Lemma~\ref{lem:kyp} can be used to turn the \(\theta\)-dependent LMIs in \eqref{eq:the1.1b} and \eqref{eq:the1.2b}  into parameter-independent ones. The dependence of the terms in 
\eqref{eq:the1.1b} and \eqref{eq:the1.2b} from \(\theta\) (which is ultimately \(\lambda_i\)) is polynomial,
so we need to define 
\begin{equation}
\phi_\nu (\theta)=\left[\theta^{\mathrm{ceil}((l+1)/2)}I_\nu,\,\,\theta^{\mathrm{ceil}((l+1)/2)-1}I_\nu,\,\,...,\,\,I_\nu  \right]^\top
\label{eq:phi}
\end{equation}
which requires
\begin{equation}
\begin{array}{cc}
\tilde{A}_\nu = U_{\nu}  \otimes I_n, &
\tilde{B}_\nu = \left[\begin{array}{c} \mathbf{0}_{(\nu-1)\times 1} \\1 \end{array}  \right] \otimes I_n,
 \\ \\
 \tilde{C}_\nu = \left[\begin{array}{c}I_{\nu} \\ \mathbf{0}_{1\times \nu}  \end{array}  \right] \otimes I_n, &
\tilde{D}_\nu = \left[\begin{array}{c} \mathbf{0}_{\nu \times 1} \\ 1 \end{array}  \right] \otimes I_n.
\end{array}
\label{eq:abcd}
\end{equation}
where  \(U_{\nu}  \in \mathbb{R}^{i{\nu}  \times {\nu} }\) is a matrix containing \(1\)'s in the first upper diagonal and \(0\)'s elsewhere, and \(\nu=n\) for \eqref{eq:the1.1b} and \(\nu=\rho-1\) for \eqref{eq:the1.2b}.
We are now ready to formulate our second main result.

\begin{the:the}
Consider \eqref{eq:formation} with given \(N\), \(A_a\), \(A_b\) and \(\mathcal{P}\) statisfying Assumption~\ref{ass:P}; excluding the first eigenvalue of \(\mathcal{P}\), which is equal to \(0\), we have that  \(\underline{\lambda}\leqslant \lambda_i\leqslant \overline{\lambda}\), with \(i=2,\, ...\,N\).
If for a chosen \(l \in \mathbb{N}\), there exist \(\tau_i \in \mathbb{R}\), and  matrices \(L_i=L_i^\top \in \mathbb{R}^{n \times n } \), and there exist \(\mathcal{D}_{\nu_k},\mathcal{G}_{\nu_k}\in\mathrm{R}^{{\nu_k}\times{\nu_k}}\), \(\mathcal{D}_{\nu_k} = \mathcal{D}_{\nu_k}^\top  \succ 0\) and \(\mathcal{G}_{\nu_k} = -\mathcal{G}_{\nu_k}^\top  \) such that
\begin{equation}
\begin{array}{c}
[\ast]^\top
M_k
\left[\begin{array}{cc} \tilde{C}_{\nu_k} & \tilde{D}_{\nu_k}
\end{array}\right]+ \\[.1cm] \ 
[\ast]^\top \!
\left[\!\!\begin{array}{cccc} -2\mathcal{D}_{\nu_k} & (\underline{\lambda}+\overline{\lambda})\mathcal{D}_{\nu_k}+\mathcal{G}_{\nu_k} \\ \ast & -2\underline{\lambda}\overline{\lambda}\mathcal{D}_{\nu_k}
\end{array}\!\!\right] \!\!
\left[\!\!\begin{array}{ccccc} I_{\nu_k} & 0 \\ \tilde{A}_{\nu_k} & \tilde{B}_{\nu_k}
\end{array}\!\!\right]\!\!\prec \!0
\label{eq:var}
\end{array}
\end{equation}
for \(k=1,2\), with \(\nu_1=n\) and \(\nu_2=\rho-1\), where \(\Gamma= [ \mathbf{0}_{n,1} \,\, I_{n}\,\, \mathbf{0}_{n,\rho-n-2} ]\) (i.e. \(\Gamma x_i = \chi_i \)), \(\Pi = [\mathbf{0}_{\rho-1,1}\,\,I_{\rho-1}  ] \), and \(\tilde{A}_{\nu_k}\), \(\tilde{B}_{\nu_k}\), 
\(\tilde{C}_{\nu_k}\), 
\(\tilde{D}_{\nu_k}\), 
\(\phi_{\nu_k}\) are defined in \eqref{eq:abcd} and  \eqref{eq:phi},
with
\begin{equation}
\phi_{\nu_1}(\lambda_i)^\top M_1 \phi_{\nu_1}(\lambda_i) =- \sum_{j=1}^l \lambda_i^j  L_j 
\label{eq:m1}
\end{equation}
\begin{equation}\begin{array}{c}
\phi_{\nu_2}(\lambda_i)^\top M_2 \phi_{\nu_2}(\lambda_i) =  \Pi (\sum_{j=1}^{\iota}{\tau_j Q_j}+ \\ \sum_{j=1}^l (\lambda_i^j (\Gamma^\top L_j  A_a + A_a^\top L_j \Gamma + \Gamma^\top L_j \Gamma ) +\\
  \lambda_i^{j+1} (\Gamma^\top L_j  A_b + A_b^\top L_j \Gamma)) \Pi^\top,\end{array}
\label{eq:m2}
\end{equation}
then
 \(\lim_{t\to\infty} ||x_i-x_j|| = 0\), \(\forall i,j \in \{1,\,...,\,N \} \).
 \label{the:var}
 \begin{proof}
 A direct application of Lemma~\ref{lem:kyp} for \(M_1\) and \(M_2\) implies that the hypotheses of Theorem~\ref{the:mai} are satisfied if the hypotheses here are.
 \end{proof}
 \end{the:the}
With this second theorem, we replace the two sets of \(N-1\) LMIs of size \(n\) and  \(\rho-1\), with only two LMIs of matrix size \(n \,\mathrm{ceil}((l+3)/2)\) and \(\rho \, \mathrm{ceil}((l+3)/2)\). This is an interesting result because the computational complexity is no longer depending on \(N\), i.e. the number of agents. On the other hand, the choice of a bigger \(l\) will improve the chances of solving the LMIs for high values of \(N\).

\section{Examples}
In order to provide a few challenging examples of application of the proposed method, we focus on a problem that is widely studied in the nonlinear dynamics community, namely the synchronisation of oscillators \cite{Pourmahmood20112853,torres2012exponential}. The approach here is of course numerical and different (or complementary) with respect to the ones found in such a literature, where the objective is usually to find a control law and then analytically prove stability. Our approach is just to propose a control law and test numerically whether it will make the subsystems converge or not.
We consider two famous examples of nonlinear systems, namely the Van der Pol oscillator \cite{grimshaw1991nonlinear} and the Lorenz attractor \cite{lorenz1963}.

\subsection{Van der Pol oscillator}
\begin{figure}[h!]
\includegraphics[width=\columnwidth]{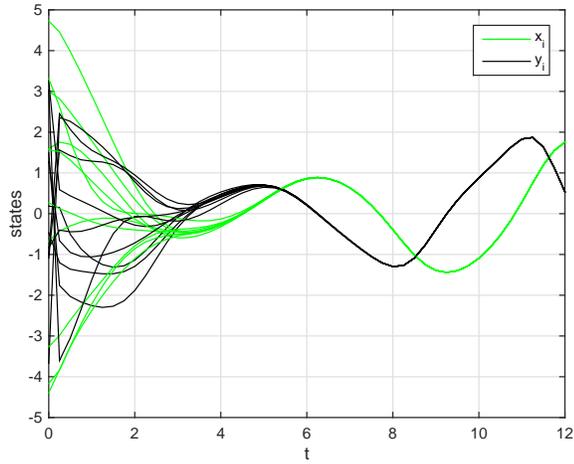}
\caption{Evolution of the state of the \(10\) coupled Van der Pol oscillators.}
\label{fig:vdp_state}
\end{figure}

\begin{figure}[h!]
\includegraphics[width=\columnwidth]{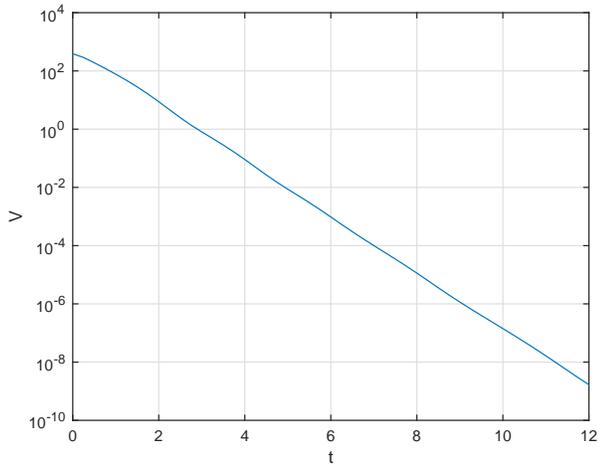}
\caption{Value of the formation Lyapunov function \(V\) for the coupled Van der Pol oscillators.}
\label{fig:vdp_lyap}
\end{figure}

We consider a system of \(N\) agents of equation
\begin{equation}
\left\{
\begin{array}{l}
\dot{x}_i = y_i\\
\dot{y}_i= \mu(1-x_i)y_i - x_i - cr_i
\end{array}
\right.
\label{eq:vanderpol}
\end{equation}
for which \(n=2\), with \(r_i= -(x_{i-1} +y_{i-1}) +2(x_{i} +y_{i})  -(x_{i+1} +y_{i+1})  \) (where the index is to be considered as modulo \(N\), i.e. \(0 \rightarrow N\), \(N+1 \rightarrow 1\)).
The interconnection between each oscillator is given by the term \(c\), a proportional feedback gain. This feedback law has just been guessed, and we use Theorem~\ref{the:var} to prove whether it works. We have coded the related LMI problem in Matlab using Yalmip \cite{yalmip}, choosing \(\mu=0.5\), \(N=10\), \(l=6\) and \(c=15\) (all arbitrary values).
By using SeDuMi \cite{sedumi} as solver, we managed to find a feasible solution, which yields a valid formation Lyapunov function. Figure~\ref{fig:vdp_state} and Figure~\ref{fig:vdp_lyap} show the evolution of the system during a simulation, with the individual states shown as well as the value of the Lyapunov function over time.

\subsection{Lorenz attractor}
\begin{figure}[h!]
\includegraphics[width=\columnwidth]{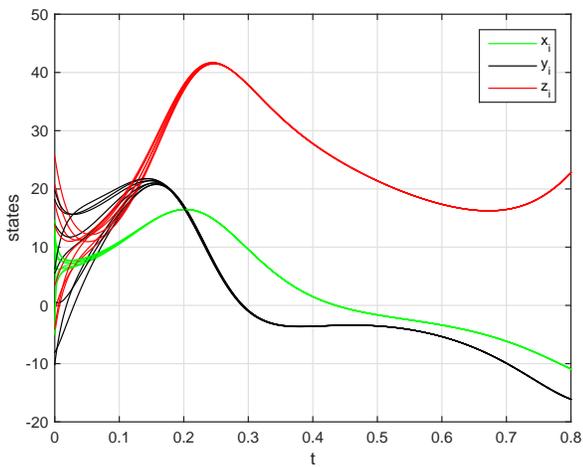}
\caption{Evolution of the state of the \(8\) coupled Lorenz systems.}
\label{fig:lor_state}
\end{figure}

\begin{figure}[h!]
\includegraphics[width=\columnwidth]{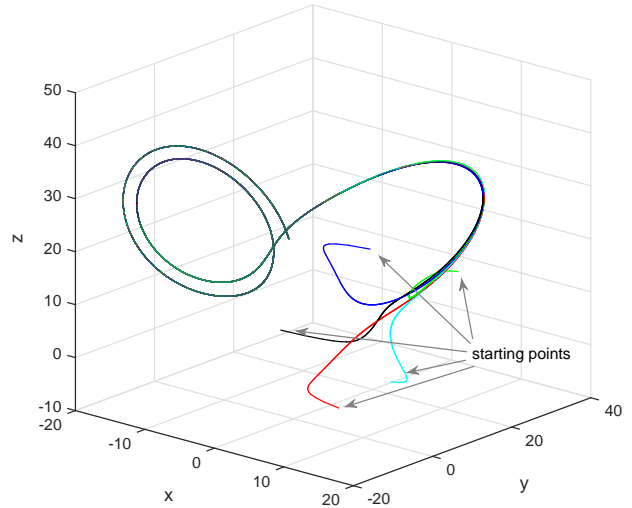}
\caption{Tridimensional visualisation of the state of some of the coupled Lorenz systems of the example (the trajectories eventually converge to the consensus trajectory).}
\label{fig:lor_3d}
\end{figure}

\begin{figure}[h!]
\includegraphics[width=\columnwidth]{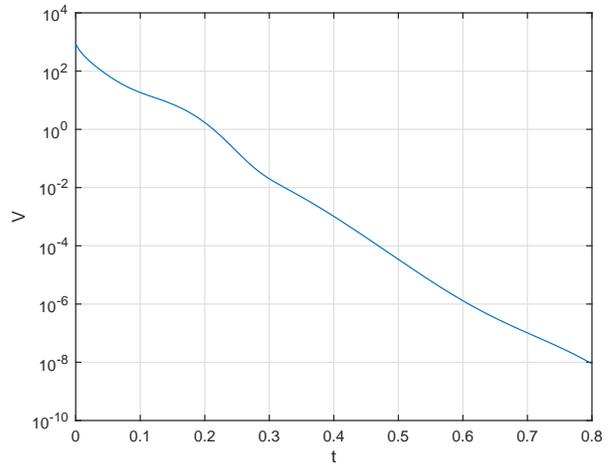}
\caption{Value of the formation Lyapunov function \(V\) for the coupled Lorenz systems.}
\label{fig:lor_lyap}
\end{figure}

We consider now system of \(N\) agents of equation
\begin{equation}
\left\{
\begin{array}{l}
\dot{x}_i = \sigma_l (y_i-x_i)-cr_{x,i}\\
\dot{y}_i = x_i(\rho_l - z_i) - y_i-cr_{y,i}\\
\dot{z}_i= x_i y_i -\beta_l z_i -cr_{z,i}
\end{array}
\right.
\label{eq:lorenz}
\end{equation}
with \(r_{i,\bullet}= -\bullet_{i-1} + 2 \bullet_{i} -\bullet_{i+1}  \) (again the index is taken modulo \(N\)). We set arbitarily \(\rho_l=28\), \(\sigma_l=10\), \(\beta_l=8/3\),  \(N=8\), \(l=6\) and \(c=50\). This time we used Theorem~\ref{the:mai}, successfully obtaining a formation Lyapunov function.
 Figure~\ref{fig:lor_state}, Figure~\ref{fig:lor_3d} and Figure~\ref{fig:lor_lyap} again show the evolution of the system during a simulation, with individual states and the value of the Lyapunov function. Notice that the Lorenz oscillator does not converge to a limit cycle but to a chaotic trajectory.

\section{Conclusion}
\label{sec:con}
We have introduced a new method for proving convergence or consensus of multi-agent system with polynomial dynamic. This method is the generalisation of the analysis methods in \cite{massioni2008distributed} and it has proven effective in test cases featuring dynamical oscillators. Further research will investigate if convex controller synthesis results can be obtained with a similar approach.




\bibliographystyle{plain}
\bibliography{article_distro}

\end{document}